\documentclass[aps,prl,twocolumn,superscriptaddress,amsmath,amssymb]{revtex4}
\usepackage{graphicx}
\usepackage{color}
\definecolor{Red}{rgb}{1.0,0.0,0.0}

\usepackage{xcolor}

\begin{document}

\title{Sublinear drag regime at mesoscopic scales in
  viscoelastic materials}

\author{A. E. O. Ferreira}
%\email{edinaldo@fisica.ufc.br}
\affiliation{Departamento de F\'{i}sica, Universidade Federal do
  Cear\'{a}, 60451-970 Fortaleza, Cear\'{a}, Brazil}

\author{J. L. B. de Ara\'ujo} 
\affiliation{Laborat\'orio de Ci\^encia de Dados e Intelig\^encia Artificial, 
Universidade de Fortaleza, 60811-905 Fortaleza, Cear\'{a}, Brazil}

 \author{W. P. Ferreira}
%\email{wandemberg@fisica.ufc.br}
\affiliation{Departamento de F\'{i}sica, Universidade Federal do
  Cear\'{a}, 60451-970 Fortaleza, Cear\'{a}, Brazil} 

\author{J. S. de Sousa}
%\email{jeanlex@fisica.ufc.br}
\affiliation{Departamento de F\'{i}sica, Universidade Federal do
  Cear\'{a}, 60451-970 Fortaleza, Cear\'{a}, Brazil}
  
\author{C. L. N. Oliveira}
\email{lucas@fisica.ufc.br}
\affiliation{Departamento de F\'{i}sica, Universidade Federal do
  Cear\'{a}, 60451-970 Fortaleza, Cear\'{a}, Brazil}

\begin{abstract}
Stressed soft materials commonly present viscoelastic signatures in
the form of power-law or exponential decay. Understanding the origins
of such rheologic behaviors is crucial to find proper technological
applications. Using an elastic network model of macromolecules
immersed in a viscous fluid, we numerically reproduce those
characteristic viscoelastic relaxations and show how the microscopic
interactions determine the rheologic response. We find that
exponential relaxations are indeed the most common behavior. However,
power laws may arise when drag forces between the macromolecules and
the fluid are sublinear, which is related to micro-deformations of the
macromolecules.
\end{abstract}

\maketitle

Purely elastic and purely viscous behaviors are limited cases of
constitutive equations of materials~\cite{Lakes2017}. Actual
substances may deform and flow, but one of these attributes usually
dominates the other, depending on the applied conditions. This
solid-liquid duality has teased researchers since at least the
19$^{th}$ century. Back then, pioneers such as James Maxwell and
Ludwig Boltzmann proposed analytical models based on series and
parallel associations of springs and dashpots to explain the peculiar
characteristics observed in silk, glass fibers, and steel
wires~\cite{Maxwell1867, Markovitz1977}. The effective response of
such early models invariably presents exponential relaxation decays,
regardless of how springs and dashpots are connected. However, these
simple approaches are only suited for some viscoelastic materials
nowadays.

In modern society, soft matter is ubiquitous and broadly
accessible. The emergence of such complex materials has triggered new
theoretical models and the improvement of proper experimental
techniques to explain and control their viscoelastic
properties~\cite{Sousa2020, Sousa2021}. Nanoindentation methods, such
as Atomic Force Microscopy, have become essential to characterize
viscoelastic features at micro and nanometer scales by probing
materials with nano-sized indenters~\cite{Sousa2017}. The
characterization of viscoelastic materials attempts to determine the
relaxation function that possesses both qualitative and quantitative
information.

Exponential and power-law relaxation functions are the two major types
of experimentally probed responses. Polyacrylamide
gels~\cite{Song2017, Calvet2004} and aqueous solutions of cationic
surfactants~\cite{Rehage1988}, for instance, present exponential-like
responses with a relaxation time for the material to achieve a new
equilibrium configuration. On the other hand, living
cells~\cite{Efremov2017}, microgel dispersions~\cite{Ketz1988}, soft
glassy materials~\cite{Sollich1998}, and hydrogels~\cite{Larson1999}
present a time-invariant power-law-like behavior.  As observed in
elastic materials~\cite{Moreira2012, Oliveira2014}, macroscopic
physical parameters are intrinsically connected to their microscopic
interactions and structures~\cite{Achar2012, Yucht2013, Milkus2017}.

Power laws and exponentials arise in many physical phenomena having a
deep origin in their dynamic processes. For instance, in non-additive
entropy systems, many physical variables are described by power-law
distributions instead of the traditional exponential functions in the
counterpart entropy~\cite{Tsallis1988, Tsallis2009}. Exponential and
power-law canonical distributions emerge naturally regarding whether
the heat capacity of the heat bath is constant or
diverges~\cite{Murilo2001}. Moreover, power laws are associated with
emergence phenomena where exponents display scaling behaviors as they
approach criticality~\cite{Stanley1971}. Systems with precisely the
same critical exponents belong to the same universality class, and a
small set of universality classes describes almost all material phase
transitions.

One of the challenges in material science is linking the physical
mechanisms at microscopic scales to macroscopic functional
behavior. This approach is especially relevant for soft matter because
properties on the molecular scale are linked to conformational and
compositional fluctuations on the nanometer and micrometer scale and,
in addition, span many orders of magnitude in
length~\cite{Praprotnik2007, Qu2011}. Soft matter holds rich
structures and various interactions at the mesoscale, where thermal
energy per unit volume is negligible, in contrast with the high energy
density stored in atomic bonds of crystalline
structures~\cite{Doi2013}. While exponential materials can be modeled
by an association of springs and dashpots, such as the so-called
standard linear solid model, power-law materials are usually obtained
by fractional rheology~\cite{West2003, Jaishankar2012} or glassy
rheology models~\cite{Fabry2001, Fabry2003}. However, these models
cannot explain the connection between macroscopic responses and their
elastic and viscous components.

We design a model of viscoelastic materials composed of an immersed
elastic network of macromolecules to study how mesoscopic interactions
influence macroscopic rheological behavior in soft
materials~\cite{Araujo2020}. We assume non-linear hydrodynamic drag
forces act between the macromolecule and the fluid, where the
contribution of elastic and viscous interactions are controlled at the
mesoscopic level. By changing the physical parameters of elastic and
drag forces, we obtain materials with exponential or power-law
relaxations or then an intermediary behavior for responses not clearly
characterized. Our results show that exponential behavior is, in fact,
the most common regime of deformation, being described by the standard
linear solid model. Power-law responses are exceptional outcomes for a
particular range of sublinear drag forces.

\begin{figure}[t]
\begin{center}
\includegraphics[width=0.9\columnwidth]{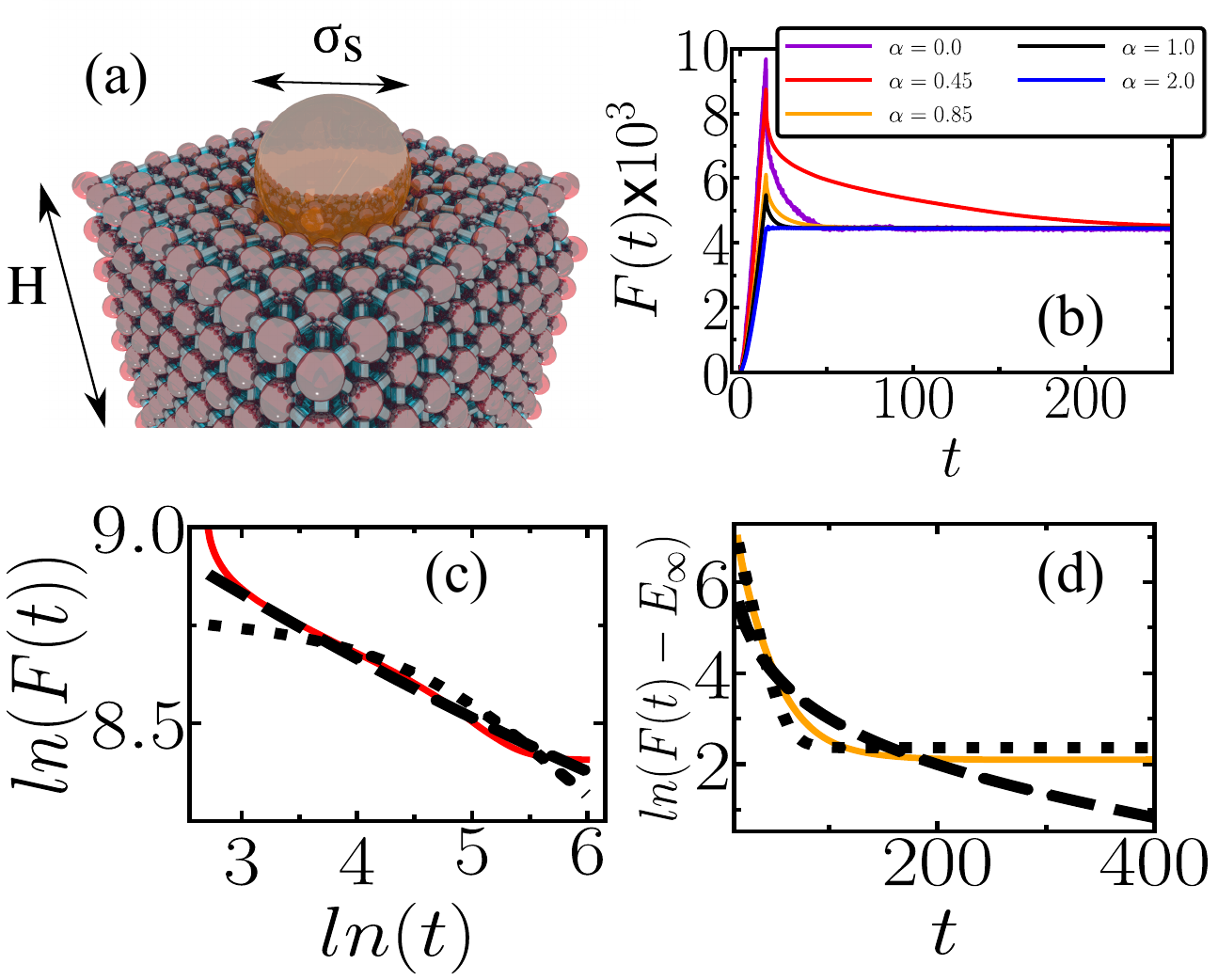}
\end{center}
\caption{(a) Viscoelastic model made by a face-centered cubic (FCC)
  lattice of macromolecules immersed in a viscous fluid (not shown in
  the image). Each spherical particle of diameter $\sigma$ represents
  a collection of molecules interacting with an effective spring
  constant $k$ with its twelve closest neighbors. A rigid spherical
  punch of diameter $\sigma_s$ indents the network from top to
  bottom. $H$ is the number of layers in the vertical direction. (b)
  Computational force curves for $\gamma=80$ and $k=800$ and several
  values of $\alpha$ exhibiting different viscoelastic
  relaxations. The cases for $\alpha=0.45$ and $\alpha=0.85$ are shown
  in panels (c) and (d), respectively, where the dashed and the dotted
  lines represent fit with the exponential and power-law model of
  Eqs.~\ref{eq:FE} and \ref{eq:FP}.}
\label{Fig1}
\end{figure}

Our model consists of $N$ spherical particles of diameter $\sigma$ and
mass $m$ arranged in a face-centered cubic (FCC) lattice with
dimensions $x\times y\times z$ given by $\sigma H\sin(\pi/3) \times
\sigma H\sin(\pi/3) \times \sigma H$, where $H$ is the number of
layers in $z$ direction, as shown in Fig.~\ref{Fig1}(a).  Every
particle interacts with its twelve nearest neighbors through an
elastic potential with an effective spring constant $k$.  The spring
network is immersed in a viscous fluid, where drag forces act on
moving particles. In this coarse-grained approach, the particles
represent macromolecules commonly found in suspended polymer chains,
colloidal aggregations, and other load-bearing structures of soft
matter.

We perform computational indentation assays to probe this network's
effective viscoelastic properties.  Firstly, a rigid spherical
indenter presses down the network at a constant rate. This
\emph{loading stage} is done during a time $\tau_l$ until a maximum
indentation depth $\delta_{max}$ is achieved. After that, called
\emph{dwell} stage, the indenter stays still while the network
rearranges towards a minimum energy configuration. Particles in the
bottom layer are not allowed to move along the $z$-axis (where
deformation is applied) but are free to slide horizontally. To avoid
finite-size effects, we limit the maximum indentation to less than
10\% of the network height~\cite{Garcia2018}, and thus we apply
$\delta_{max}\approx \sigma$.

The equation of motion of the $i_{th}$ particle, at position $\vec
r_i$, is given by the following
equation~\cite{Langevin1908,Lemons1997}
\begin{equation}
 m\frac{d^2 \vec r_i}{dt^2} = - \nabla U_i - \gamma v^{\alpha}_i\hat v_i,
\label{eq:dmj}
\end{equation}
where $U_i$ is the interaction potential of particle $i$ due to other
particles and the indenter, given by
\begin{equation}
 U_i = \frac{k}{2} \sum_j (r_{ij}-\ell)^2 + \epsilon
 \left[\frac{\sigma}{r_{is}-(\sigma_s-\sigma)/2}\right]^{\xi}.
\label{eq:U}
\end{equation}
The summation in the first part runs over the neighbors, where
$r_{ij}=|\vec{r}_i-\vec{r}_j|$ and $\ell$ are, respectively, the
distance and the equilibrium distance between the centers of particles
$i$ and $j$. The last term in Eq.~(\ref{eq:U}) represents a hard-core
potential applied only to those particles in contact with the
indenter, where $\epsilon$ is an energy parameter, and
$r_{is}=|\vec{r}_i-\vec{r}_s|$ is the distance between the center of
the particle and the indenter. The exponent $\xi$ must be large enough
to keep the stiffness of the indenter.

The last term of Eq.~(\ref{eq:dmj}) represents a generalized drag
force acting oppositely to the particle velocity, $\vec v_i = v_i\hat
v_i$, with magnitude given by $\gamma v^{\alpha}_i$, where $\gamma$
and $\alpha$ are related to the particle geometry and the fluid
properties in which the particles are immersed. Dissipation vanishes
for $\gamma=0$, leading to purely elastic networks where our model
reproduces the well-known Hertz behavior for mechanical
contacts~\cite{Araujo2020}. However, when local friction becomes
relevant, $\gamma > 0$, the model may produce distinguished behaviors
of viscoelastic materials. Notice that $\alpha = 1$ and $\alpha = 2$
represent typical values for drag forces acting on rigid
structures. The linear regime, known as Stoke's law, arises for small
Reynolds numbers when viscous forces dominate over inertial forces,
where $\gamma$ is proportional to the medium's viscosity and the
particle's diameter. On the other hand, the quadratic drag is dominant
for large Reynolds numbers. In this case, $\gamma$ is proportional to
the medium's density and the cross-sectional area of the particle.

\begin{figure*}[t!]
\begin{center}
\includegraphics[width=0.9\textwidth]{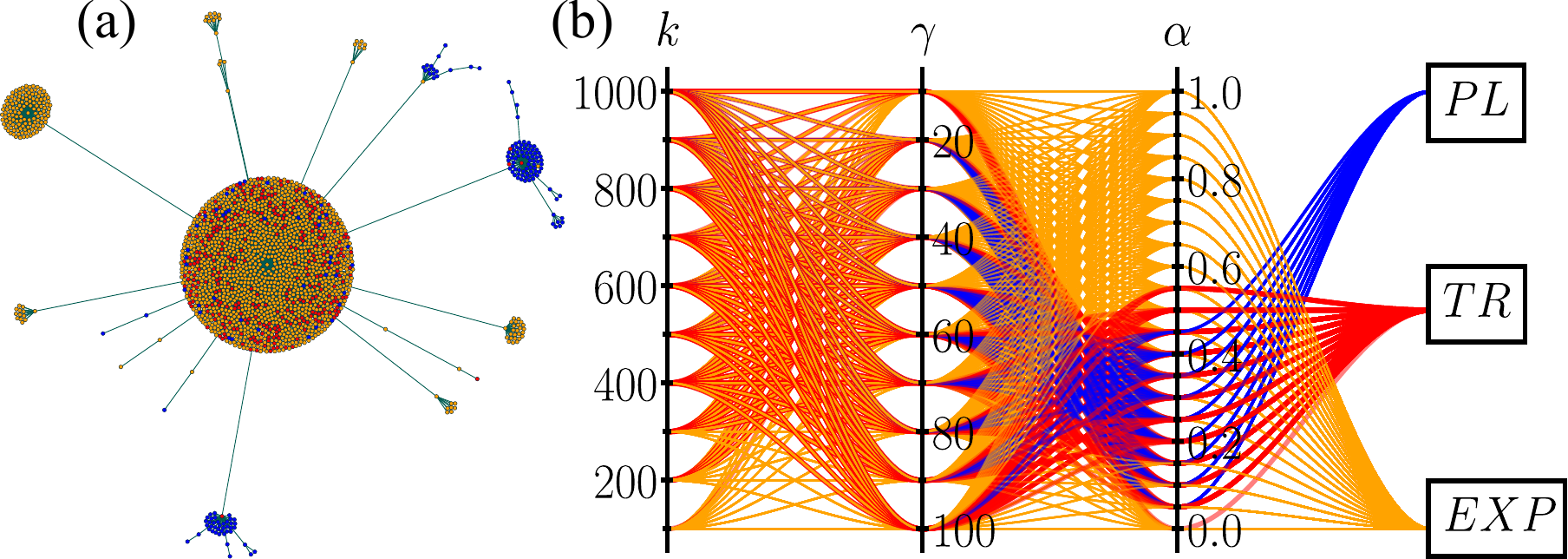}
\end{center}
\caption{(a) Graph visualization of the clustering process using
  \emph{K-means} method. Each dot represents a computational
  experiment for a given combination of $k$, $\gamma$, and $\alpha$,
  and the colors represent different relaxation outcomes, namely,
  power-law (PL) in blue, exponential (EXP) in yellow and transitional
  behavior (TR) in red. (b) Parallel plot showing how different values
  of the mesoscopic network parameters lead to different types of
  viscoelastic relaxation. Each line crosses a combination of $k$,
  $\gamma$, $\alpha$, and the corresponding rheological behavior. }
\label{Fig2}
\end{figure*}

The physical origins of the sublinear regime ($\alpha<1$), however,
are entirely different and are related to the
deformability/adaptability of bodies subjected to drag
forces~\cite{Vogel1994}. Many living beings present sublinear drag
behaviors, where deformability is a survival strategy to protect their
fragile structures under hydrodynamic conditions. Characterizing the
drag exponents in deformable systems is recurrent in botany,
aerodynamics, and hydrodynamics~\cite{Vogel1989, Favier2009,
  Gosselin2010, John2015}. Typical drag exponents for algae are as
small as 0.34~\cite{Vogel1984}. On the other hand, tulip and willow
oak trees are more rigid and present exponents close to unity,
$\alpha=0.92$ and $\alpha=0.94$~\cite{Vogel1989}, respectively. The
limiting case of $\alpha=0$ corresponds to constant frictional forces,
regardless of the particle velocity. In our model, micro deformations
are associated with a new molecular arrangement of the macromolecules.

The numerical solution of Eq.~(\ref{eq:dmj}) is performed through
molecular dynamics simulations~\cite{Rapaport2004, Araujo2017} with
periodic boundary conditions applied to the horizontal plane, and time
integration is done with the velocity Verlet algorithm with time step
$dt$. See~\cite{Param} for the parameters used. The force is computed
as the sum of all collisions on the indenter at each time step. Each
set of $k$, $\gamma$, and $\alpha$ represents a specific material and
defines the macroscopic viscoelastic responses. To investigate how
microscopic properties lead to the rheological behavior of the entire
network, we perform simulations varying the spring constant between
100 and 1000, the drag constant between 10 and 100, and the drag
exponent between 0.0 and 1.0, totalizing 2100 different networks.

Figure~\ref{Fig1}(b) shows typical force curves for $\gamma=80$ and
$k=800$ and different values of the drag exponent. $F(t)$ is the
computational analogous of an indentation assay used to characterize
macroscopic responses of actual materials. In the loading stage, the
indenter slowly deforms the network until the contact force reaches
its maximum value, which is inversely proportional to $\alpha$. In the
dwell stage, when the indenter stays still, $F(t)$ relaxes until part
of the mechanical energy is lost through friction.  The timescale of
such dissipation depends on $\alpha$ in a highly complex fashion. For
$\alpha = 2$, dissipation is fast enough so that the response behavior
is qualitatively identical to a perfectly elastic network. For
$\alpha<1$, drag forces slowly remove energy from the network leading
to a viscoelastic decay.

In a continuum approach, the analytical time-dependent contact force
of a viscoelastic sample indented by a spherical punch follows the
convolution integral~\cite{Araujo2020}
\begin{equation}
    \mathcal{F}(t) = \int_0^t
    R(t-t^{'})\frac{d\delta^{\frac{3}{2}}(t^{'})}{dt^{'}} dt^{'},
        \label{eq:forceint}
\end{equation}
where $R(t)$ is the relaxation function and $\delta(t)$ the
indentation depth history. The contact force in
Eq.~(\ref{eq:forceint}) is normalized by the constant
$4\sqrt{\sigma_s}\delta_{max}^{\frac{3}{2}} / 3(1-\nu^2)$, where $\nu$
is the Poisson coefficient. Typical relaxation behaviors for
viscoelastic materials are given by
\begin{eqnarray}
  R_P(t) &=& E_{\infty} + \Delta E \,\, t^{-\beta},\nonumber\\
  R_E(t) &=& E_{\infty} + \Delta E \,\, e^{-\frac{t}{\tau}},
    \label{eq:relax}
\end{eqnarray}
where $R_P(t)$ is the relaxation model for power-law materials with an
exponent $\beta$ and $R_E(t)$ for exponential materials with
relaxation time $\tau$. In both models, $E_{\infty}$ is the elastic
modulus at considerable times when the material is completely relaxed,
and $\Delta E$ is the difference between the maximum value, at
$t=\tau_l$, and $E_{\infty}$. We assume that materials become
perfectly elastic after some time, although this long-time elasticity
plateau is only sometimes observed in actual
materials~\cite{Footnote}.  However, micro-deformations are allowed
only due to the movement of the pseudo-atoms, which is represented by
the sublinear drag regime. Solving Eq.~(\ref{eq:forceint}) with
Eqs.~(\ref{eq:relax}) in the dwell stage leads, respectively, to
\begin{eqnarray}
    \label{eq:FE}
    \mathcal{F}_P(t) &=& E_\infty + a \, t^{-\beta},\\
    \mathcal{F}_E(t) &=& E_\infty + b \, e^{-\frac{(t-\tau_l)}{\tau}} - c \, e^{-\frac{t}{\tau}},
    \label{eq:FP}
\end{eqnarray}
where the constant $a\approx \frac{\Delta E}{\Gamma(1-\beta)}$ is
obtained by expanding the incomplete beta function, $\Gamma(1-\beta)$
is the gamma function, and $b=\frac{3\tau}{2\tau_l} \Delta E$ and
$c=\frac{3\sqrt{\pi}}{4}(\frac{\tau}{\tau_l})^{\frac{3} {2}}\Delta E
\, \mathrm{erfi}(\sqrt{\frac{\tau_l}{\tau}})$, where
erfi($\sqrt{\frac{\tau_l}{\tau}}$) is the imaginary error function of
$x$.

Fitting the dwell part of the force curve allows us to map the
macroscopic mechanical properties ($E_\infty$, $\Delta E$, $\tau$,
$\beta$) with the mesoscopic parameters ($k$, $\gamma$ and $\alpha$)
as done previously for $\alpha=1$~\cite{Araujo2020}. Here we focus on
finding the qualitative rheological behavior rather than describing
relations among parameters. Figures~\ref{Fig1}(c) and (d) show the
same force curves as in (b) for $\alpha=0.45$ and $\alpha=0.85$,
respectively, where the curves are fitted both with $\mathcal{F}_P(t)$
and $\mathcal{F}_E(t)$. Clearly, the $\alpha=0.45$ case is better
fitted with the power-law model, while $\alpha=0.85$ with an
exponential.

The mean-square error determines the goodness-of-fit parameter between
the obtained numerical force curve $F(t)$ and the analytical force
models given in Eqs.~\ref{eq:FE} and \ref{eq:FP},
\begin{equation}
  \chi_M = \frac{1}{N_p}\sum_{i=1}^{N_p} \left[F(t_i)-\mathcal{F}_M(t_i)\right]^2,
\end{equation}
where $N_p$ is the number of points in the force curves and $M$ stands
for exponential ($E$) or power-law ($P$) model type. This statistical
index, calculated for each combination of $k$, $\gamma$, $\alpha$, is
used in a \emph{K-means} method as an unsupervised clustering
strategy~\cite{Geron2019, Sinaga2020} to classify the computational
force curves as either an exponential or a power-law behavior, or even
a transitional regime that cannot be clearly classified. In this
machine learning process, the classification of the material takes
into account not only the values of $\chi_{E}$ and $\chi_{P}$ but also
the trends and distributions in the phase space of $\chi_{E}$ and
$\chi_{P}$. Figure~\ref{Fig2}(a) shows the graph visualization of this
clustering process. The exponential behavior is indeed the most common
one, as shown in the big cluster in Fig.~\ref{Fig2}(a), but small
groups of power-law materials are also observed.  The parallel
coordinates plot in Fig.~\ref{Fig2}(b) summarizes the rheological
outcomes. Each line passes through every combination of $k$, $\gamma$,
and $\alpha$ ending up in one of the three boxes representing the
response behavior of the network.

\begin{figure}[t]
\begin{center}
\includegraphics[width=0.9\columnwidth]{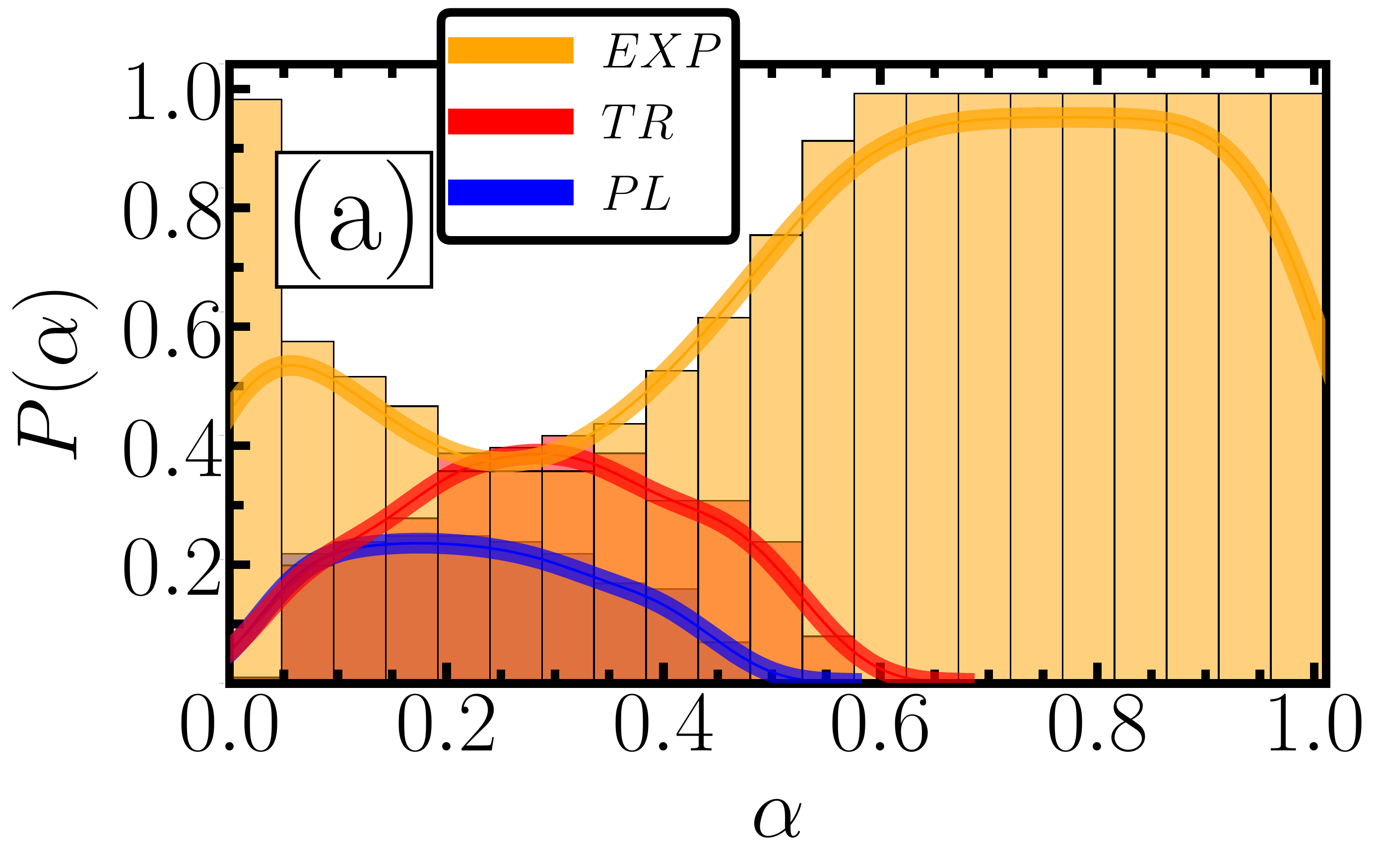}
\includegraphics[width=0.9\columnwidth]{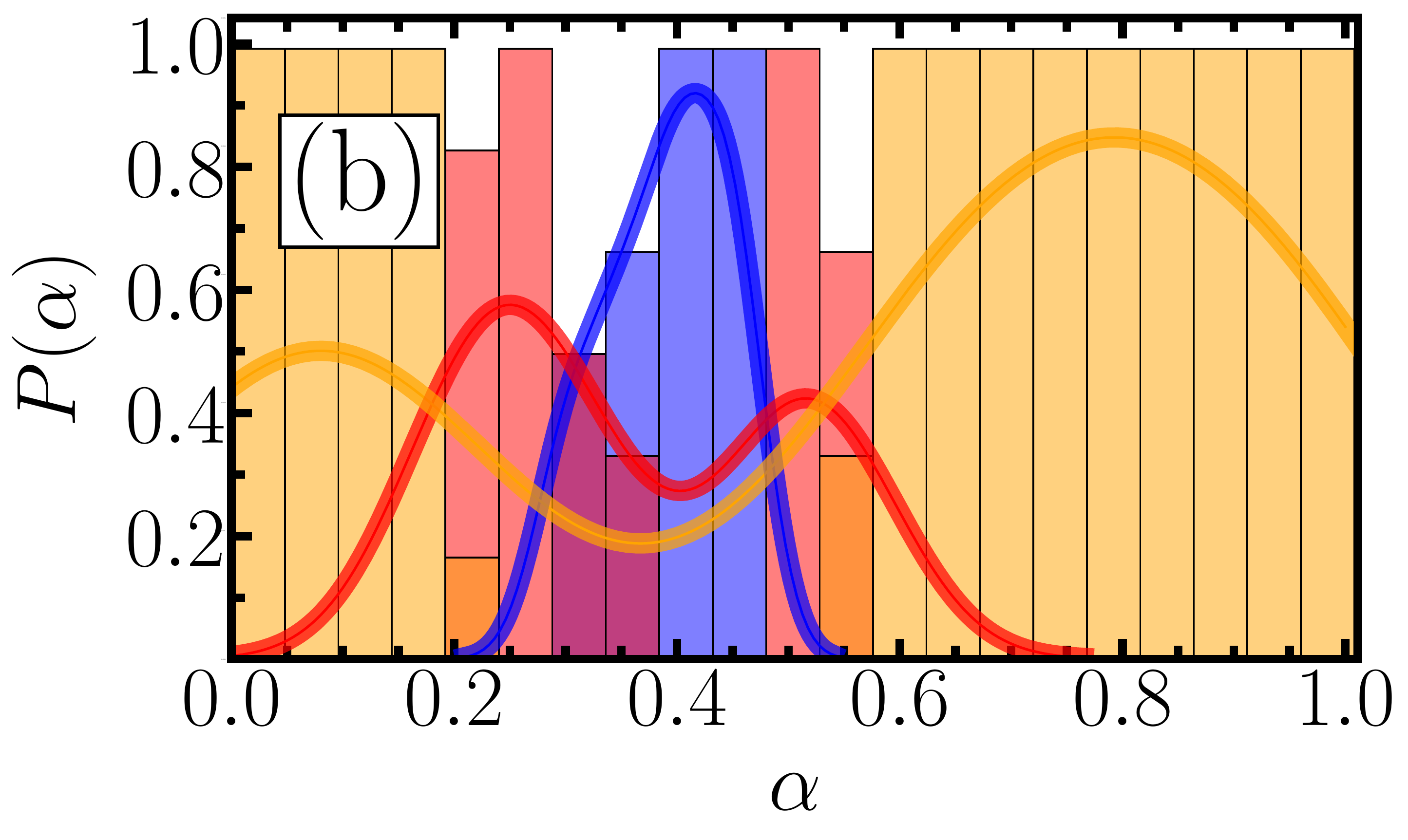}
\end{center}
\caption{The probability that a given $\alpha$ leads to a power law
  (blue bars), an exponential (yellow bars), or a transitional
  response (red bars). The solid lines show the probability density
  computed by the KDE (kernel density estimation) algorithm. Panel (a)
  shows results for the whole range of $k$ and $\gamma$, while panel
  (b) considers only networks with $k$ ($k\geq 800$) and $\gamma$
  ($\gamma\geq 80$).}
\label{Fig3}
\end{figure}

To understand why small clusters of power-law materials form in
Fig.~\ref{Fig2}(a), we must investigate the impact of considering
sublinear drag regimes. Fig.~\ref{Fig3} shows the normalized
probability $P(\alpha)$ of finding PL, EXP, or TR behavior for a given
$\alpha$. In panel (a), $P(\alpha)$ is computed for all values of $k$
and $\gamma$ considered here. The three distributions strongly
overlap, making classification a challenging exercise.  In panel (b),
on the other hand, we remove small values of $k$ and $\gamma$. For
networks not too soft, $k \geq 800$, and not too elastic, $\gamma \geq
80$, those distributions split for different regions of $\alpha$, and
the drag exponent becomes the central controller to characterize the
macroscopic behavior. Exponential behaviors are found mainly for
$\alpha \lessapprox 0.2$ and $\alpha \gtrapprox 0.55$, while the
relaxation is a power law for $\alpha$ between 0.3 and 0.45.  For
intersecting values of $\alpha$, the superposition of the
probabilities of PL and EXP leaves doubt in classifying the material
either as an exponential or a power law.

\begin{figure}[t]
\begin{center}
\includegraphics[width=0.9\columnwidth]{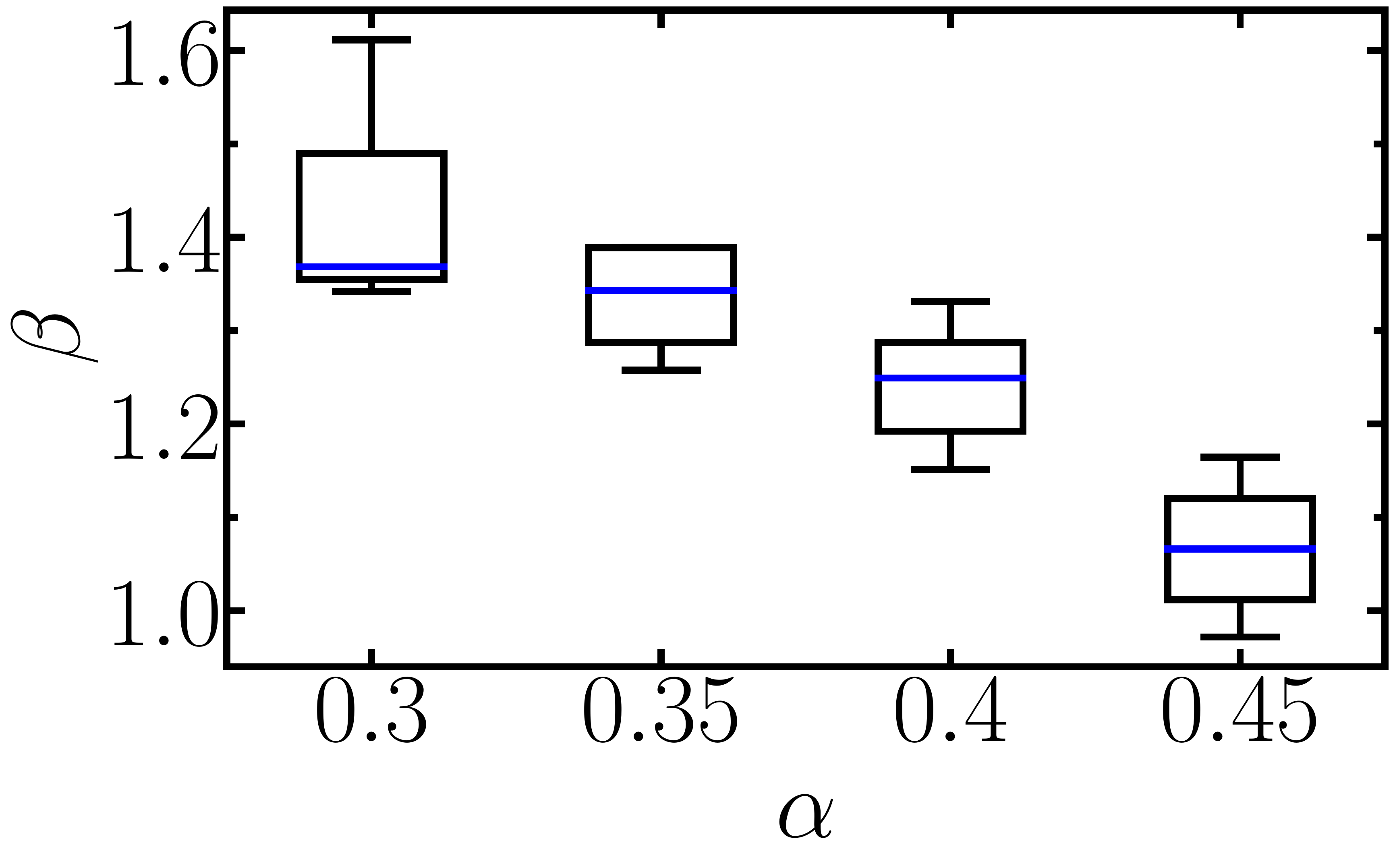}
\end{center}
\caption{For those networks classified as a power-law material in
  Figure~\ref{Fig2}(b), we show the relationship between the
  relaxation exponent $\beta$ and the mesoscopic drag exponent
  $\gamma$.}
\label{Fig4}
\end{figure}

For those cases where the network presents a power-law behavior, we
show in Fig.~\ref{Fig4} the relationship between the macroscopic
relaxation exponent $\beta$ and the drag exponent. For each value of
$\alpha$, there is a small dispersion distribution of $\beta$ whose
mean range is between 1.05 and 1.35. Actual materials exhibiting
power-law relaxation usually present smaller exponents and exhibit
structural disorder and metastability~\cite{Sollich1997, Fabry2001,
  Fabry2003, Jaishankar2012}. Computational investigations of
disordered two-dimensional networks obtained relaxation exponents
between 0.5 and 0.75, depending on the network
arrangement~\cite{Milkus2017}.  Our simulations exhibit macroscopic
relaxation exponents above 1.0, mainly because our viscoelastic solid
model is structurally ordered and stable, restricting the viscoelastic
responses to faster relaxation regimes. Our results clearly show that
structural disorder is not a mandatory ingredient for power-law-like
viscoelasticity and can only change the exponent.

In conclusion, we show why viscoelastic materials present power-law or
exponential relaxation responses. Using molecular dynamics
simulations, we perform numerical indentations onto a network of
interacting macromolecules immersed in a fluid and reproduce typical
viscoelastic signatures as those found experimentally. The
macromolecules interact with the fluid through a non-linear drag
regime given by $\gamma v^{\alpha}$, where $\gamma$ and $\alpha$ are
mesoscopic parameters, and $v$ is the particle's velocity.  For each
set of the interacting parameters, we classify the macroscopic
viscoelasticity using an unsupervised clustering algorithm according
to the type of relaxation of the deformed network, namely, exponential
or power-law relaxations or transitional behavior between them. While
exponential behaviors are predominant, power laws may arise in the
sublinear regime. In fact, the drag exponent alone may explain the
macroscopic viscoelastic relaxation for materials not too elastic nor
too soft. More specifically, power-law responses are found for
$0.3\lessapprox \alpha \lessapprox 0.45$, while exponential responses
for $0.0\lessapprox \alpha \lessapprox 0.2$ and $0.55\lessapprox
\alpha \lessapprox 1.0$.

\begin{acknowledgments}
The authors acknowledge the financial support from the Brazilian
agencies CNPq, CAPES, and FUNCAP.
\end{acknowledgments}


\begin{thebibliography}{100}
\bibitem{Param} The dimensionless length, mass, and energy units are
  given, respectively, by $\sigma=1$, $m=1$, $\epsilon=1$, and the
  timestep $dt=0.001t_0$, where the time unit has the following
  relation $t_0=\sqrt{\frac{m\sigma^{2}}{\epsilon}}$.  Other length
  parameters are given by $\sigma_s=10\sigma$, $\ell=\sigma$, and
  $H=15\sigma$, and some exponents by $\nu=0.5$ and $\xi=400$. The
  number of particles in the network is $N=2748$.
\bibitem{Footnote} This assumption is necessary to avoid high-pressure
  effects in our numerical model. For instance, if the external force
  is large enough to ultimately compress the small particles until a
  condensate network is formed, any additional pressure would induce
  the deformation of those particles.
\end{thebibliography}
\end{document}